\documentclass[secnumarabic,amssymb, nobibnotes, aps,prl,twocolumn,longbibliography]{revtex4-1}

\usepackage{amsmath}
\usepackage{graphicx}
\usepackage{mathtools}
\usepackage{xcolor}
\def \basp{BaFe$_2$(As$_{1-x}$P$_x$)$_2$}

\usepackage[english]{babel}

\begin{document}
\title{Magnetoresistance scaling, disorder, `hot spots' and the origin of $T$-linear resistivity in \basp}

\author{Nikola Maksimovic}
\altaffiliation{Contact for correspondence, nikola\_maksimovic@berkeley.edu}
\affiliation{Department of Physics, University of California, Berkeley, California 94720, USA}
\affiliation{Materials Science Division, Lawrence Berkeley National Laboratory, Berkeley, California 94720, USA}

\author{Ian M. Hayes}
\affiliation{Department of Physics, University of California, Berkeley, California 94720, USA}
\affiliation{Materials Science Division, Lawrence Berkeley National Laboratory, Berkeley, California 94720, USA}

\author{Vikram Nagarajan}
\affiliation{Department of Physics, University of California, Berkeley, California 94720, USA}
\affiliation{Materials Science Division, Lawrence Berkeley National Laboratory, Berkeley, California 94720, USA}

\author{Alexei E. Koshelev}
\affiliation{Materials Science Division, Argonne National Laboratory, Lemont, Illinois 60439, USA}

\author{John Singleton}
\affiliation{National High Magnetic Field Laboratory, Los Alamos National Laboratory, Los Alamos, New Mexico 87545, USA}

\author{Yeonbae Lee}
\affiliation{Materials Science Division, Lawrence Berkeley National Laboratory, Berkeley, California 94720, USA}

\author{Thomas Schenkel}
\affiliation{Materials Science Division, Lawrence Berkeley National Laboratory, Berkeley, California 94720, USA}

\author{James G. Analytis}
\altaffiliation{Contact for correspondence, analytis@berkeley.edu}
\affiliation{Department of Physics, University of California, Berkeley, California 94720, USA}
\affiliation{Materials Science Division, Lawrence Berkeley National Laboratory, Berkeley, California 94720, USA}

\begin{abstract}
The scaling of $H$-linear magnetoresistance in field and temperature was measured in under-doped (x = 0.19) and optimally-doped (x=0.31)~\basp. We analyze the data based on an orbital model in the presence of strongly anisotropic quasiparticle spectra and scattering time due to antiferromagnetism. The magnetoresistance is dominated by the properties of small regions of the Fermi surface called `hot spots' where antiferromagnetic excitations induce a large quasiparticle scattering rate. Approximate temperature-magnetic field scaling relations are derived and shown to be consistent with the experimental data. We argue that these results link the origin of linear-in-temperature resistivity to hot spots arising from an antiferromagnetic critical point, and magnetoresistance measurements provide a route to quantify this link.
\end{abstract}

\maketitle

\section{Introduction}
The defining signature of a strange metal is a resistivity that varies linearly with temperature, and seems to cross intrinsic energy scales (e.g. the Debye temperature and Mott-Ioffe-Regel limit) with impunity --- it is thought that this behavior stems from quantum critical physics, though an agreed upon mechanism is still not established. In recent years, the magnetoresistance of quantum critical metals has become a subject of intense study, providing another avenue to probe their properties. In particular, in typical metals, the magnetoresistance (MR) varies quadratically with field and is determined by a combination of temperature-dependent and temperature-independent contributions to the resistivity~\cite{Pippard2009}. By contrast, in many quantum critical metals, the MR has been observed to vary linearly with magnetic field~\cite{Hayes2016,Sarkar2019,Prozorov,Licciardello2019,Niu2017,Nakajima2019,Sales2016,Chu2019,Kumar,Giraldo-Gallo2018}, and scale only with the temperature-dependent resistivity, suggesting a non-trivial connection between magnetic field ($H$) and temperature ($T$) in such materials.

To understand the nature of the scaling it is useful to partition the resistivity into two contributions --- a temperature-independent contribution $\rho_0$, typically arising from scattering from defects, and a temperature-dependent contribution $\rho_t$ that may arise from charge carrier interactions with phonons, quasiparticle excitations, order parameter fluctuations, and so on. In the case of strange metals, it has been experimentally established that $\rho_t\approx\alpha k_BT$, with $\alpha$ some phenomenological constant of proportionality.

The magnetoresistance scaling arises because the interplay of field and temperature can captured by a quadrature sum, which when rearranged has the following form
\begin{equation}
    {\rho(H) - \rho_0\over \alpha k_BT} \sim \sqrt{1+ { \left(\eta H\over \alpha k_BT\right)^2}}, 
    \label{eq:qcmr}
\end{equation}
where $\rho(H)$ is the field dependent resistivity at temperature $T$, and $\eta$ a parameter that plays a similar role for the field dependence as $\alpha$ does for the temperature dependence~\cite{Hayes2016}. Eq. \ref{eq:qcmr} was phenomenologically motivated by measurements of~\basp~near its antiferromagnetic quantum critical point. Since then, a growing number of putative quantum critical metals have shown qualitatively similar behavior~\cite{Hayes2016,Sarkar2019,Prozorov,Licciardello2019,Niu2017,Nakajima2019,Sales2016,Chu2019,Kumar,Giraldo-Gallo2018}, albeit with notable deviations in the quantities $\alpha$ and $\eta$. The observation of $H$-linear magnetoresistance is unusual but not unprecedented. There are multiple possible causes of this including the presence of Dirac quasiparticles~\cite{Abrikosov2000,Pal2013}, sample heterogeneity~\cite{majumdar_dependence_1998, Ramakrishnan2017}, guiding center diffusion in a smooth random potential~\cite{Song2015}, fluctuations from spin density waves~\cite{Rosch1999m,Koshelev2016}, or singular regions of the Fermi surface where the Fermi velocity changes discontinuously~\cite{Pippard2009,Koshelev2013}. The scaling in Eq. (1), however, is surprising because it conflicts with Kohler's rule for orbital magnetoresistance~\cite{Pippard2009}. The mathematical statement of which is
\begin{equation}
    {\rho(H)-\rho(0)\over\rho(0)} = f\left({H\over\rho(0)}\right),
    \label{eq:kohler}
\end{equation}
where $f$ is a smooth and usually positive function. The crucial difference between Eq.~\ref{eq:kohler} and Eq.~\ref{eq:qcmr} is that the denominator in Eq.~\ref{eq:kohler} involves both temperature-dependent and temperature-independent contributions $\rho(0) = \rho_0 + \rho_t$, and not just $\rho_t = \alpha k_BT$. Kohler scaling applies in cases of a single, uniform momentum relaxation rate. The absence of disorder scattering, $\rho_0$, in Eq.~\ref{eq:qcmr} suggests that the relevant scattering determining the magnetoresitance is only the $T$-linear component. It is in this sense that the $H$-linear resistivity is thought to probe the same physics as $T$-linear resistivity in quantum critical metals. This immediately begs the question of what the role of disorder is and why, at first glance, it does not appear to affect the magnetoresistance. It is the purpose of this study to examine this question systematically.

In this study we show that $H$-linear magnetoresistance and its scaling with the temperature-dependent component of the resistivity is not peculiar to compositions of \basp~in the quantum critical regime, but also describes the magnetoresistance at moderate P-substitution levels in the antiferromagnetically ordered phase. We show that this is consistent with a physically reasonable model of magnetoresistance dominated by singular regions of the Fermi surface (`turning points') crossing the Bragg planes of the broken symmetry state~\cite{Koshelev2013}. The data in the quantum critical regime is also consistent with such a model, where Bragg points become `hot spots' characterized by rapid quasiparticle scattering from spin fluctuations~\cite{Koshelev2016,Rosch1999m}. These models reproduce the $H$-linear magnetoresistance, and suggest that the quasiparticle scattering rate varies around the Fermi surface between `hot' and `cold' parts. The data suggest that scattering from spin fluctuations dominates at the hot spots and consequently controls the magnetoresistance. This leads to approximate `$H-T$' scaling relations, which hold over wide ranges of field and temperature both in the antiferromagnetic and quantum critical regimes. A natural consequence of this result is that the $T$-linear zero-field resistivity of the strange metal is connected to the properties of the `hot' spots.  The limitations of these models are left to the discussion.

\section{Results}
\subsection{Antiferromagnetically ordered regime, \basp~with x = 0.19}
\begin{figure*}[!htpb]
\centering
\includegraphics[scale=0.75]{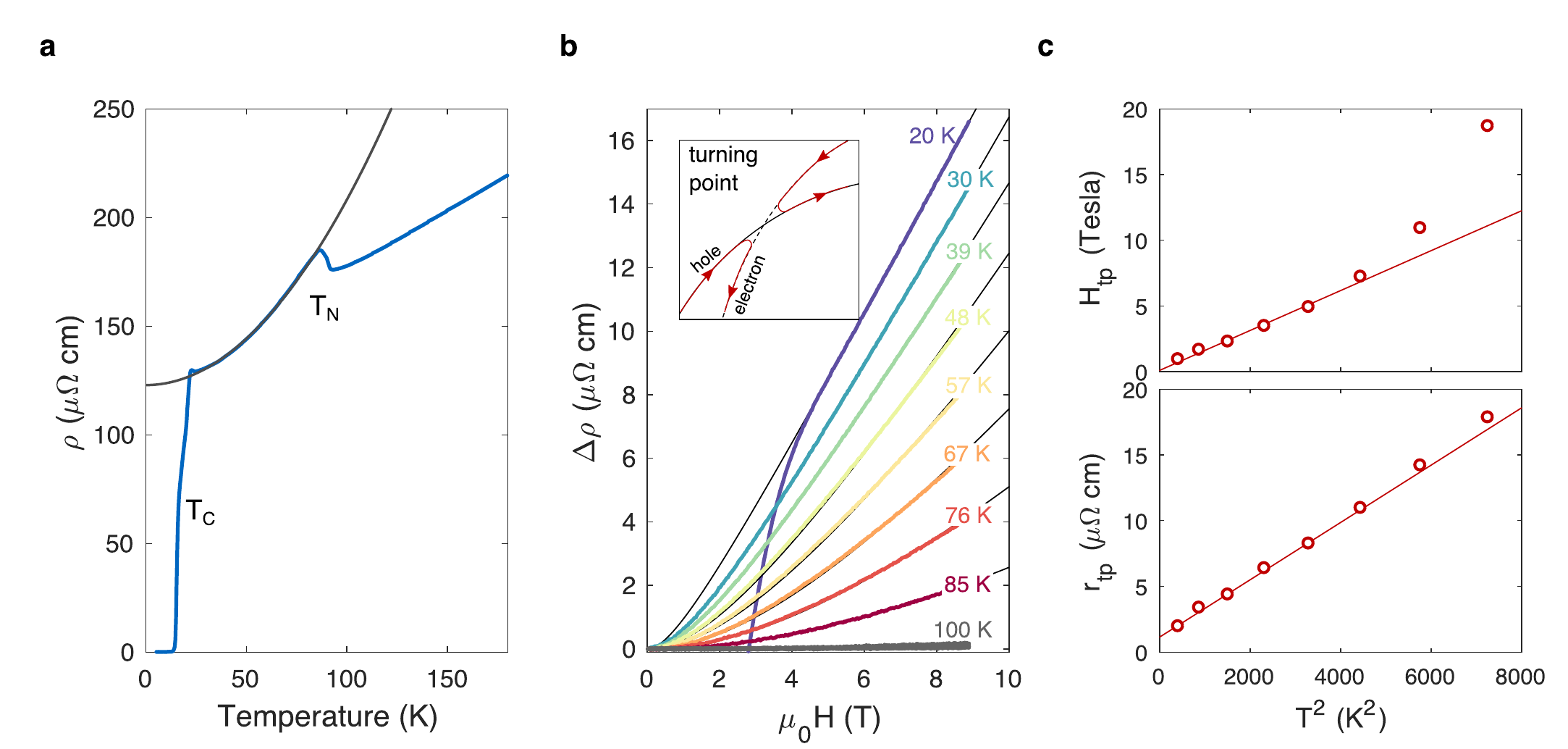}
\caption{{\bf Transport in BaFe$_2$(As$_{0.81}$P$_{0.19}$)$_2$ and magnetoresistance model based on turning points} (a) Resistivity shows a transition to an AFM ordered state ($T_{N} \approx 95 K$). Inside the AFM state, the resistivity varies with $T^{2}$, with a finite $T = 0$ intercept. The data are fitted well by $\rho(H=0) = 122.8 [\mu\Omega cm] + 0.0085[\mu\Omega cm / K^{2}] \times T^{2}$ (black line). (b) Magnetoresistance for different temperatures with fits to the turning point MR model (Eq.~\ref{eq:tp}) indicated by black lines. The inset shows a cartoon illustration of the quasiparticle orbits near the turning point in the AFM state. (c) The fit parameters of the model are plotted as a function of temperature squared, with a best fit line to the data below 70 K. $H_{tp} = 0.098 [T] + 0.0015 [T/K^{2}] \times T^{2}$, and $r_{tp} = 1.1 [\mu\Omega cm] + 0.0022 [\mu\Omega cm / K^{2}]\times T^{2}$.}
\label{fig:underdoped}
\end{figure*}

In order to elucidate the source of scaling in the quantum critical regime, it is useful to examine the magnetoresistance at a lower doping level inside the antiferromagnetically (AFM) ordered state. Here, we examine a single crystal of ~\basp~with $x=0.19$, where the antiferromagnetic N\'eel transition temperature is $T_{N} \approx 95$K. In Fig.~\ref{fig:underdoped}, transport data are shown for this crystal. Fig.~\ref{fig:underdoped}a shows that the resistivity at zero applied field varies with $T^{2}$ over a broad range of temperature inside the AFM ordered state with a finite intercept at $T = 0$. At this composition, the resistivity is likely influenced by quasiparticle scattering from diffuse spin excitations~\cite{Volkenshtein1973,Tucker2014}.

Fig.~\ref{fig:underdoped}b shows that the MR inside the AFM state has a hyperbola-like shape, approaching a linear dependence on $H$ at high fields. The magnetoresistance gets broadened and damped as temperature increases. A model for this behavior was developed in Ref.~\cite{Koshelev2013}, which is illustrated in the inset of Fig.~\ref{fig:underdoped}b. AFM order opens a gap at points on the Fermi surface nested by the AFM ordering vector. As a quasiparticle undergoes orbital motion in a magnetic field, the Fermi velocity is rapidly reversed at `turning points' due to the AFM coupling between the electron-like and hole-like pockets. This mechanism produces an $H^{2}$ variation of the MR at low fields, which crosses over to linear variation at higher fields as the number of quasiparticles pushed through the turning point by the Lorentz force increases with field~\cite{Koshelev2013,Pippard2009}. The turning points are expected to dominate the overall MR because of their large Fermi surface curvature~\cite{Koshelev2013}. This is consistent with the data in Fig.~\ref{fig:underdoped}b, where the MR rises sharply as temperature decreases below the AFM ordering temperature. Thus, we neglect the potential MR contribution from the `cold' parts of the Fermi surface away from the turning point. Here, we summarize the result of this model, which is described in more detail in the supplemental material
\begin{equation}
    \rho(H) - \rho(0) = r_{tp} \mathcal{B}\left(H/H_{tp}\right)   \label{eq:tp}.
\end{equation}
$H_{tp} \propto \Delta_{tp}/\tau_{tp}$ and $r_{tp} \propto \Delta_{tp}/\tau_{tp}$. $\Delta_{tp}$ is the size of the AFM gap and $\tau_{tp}$ is the quasiparticle scattering rate in the vicinity of the turning point. $\mathcal{B}$ is a dimensionless mathematical function that varies with $H^{2}$ when $H<H_{tp}$ and crosses over to $H$-linear when $H>H_{tp}$. In Fig.~\ref{fig:underdoped}b, we find that the data are well fitted by Eq.~\ref{eq:tp} with the temperature-dependent parameters shown in Fig.~\ref{fig:underdoped}c.

Here, we use this model to derive a simple $H-T$ scaling relation. In Eq.~\ref{eq:tp}, the dimensionless function $\mathcal{B}$ determining the field-dependence of the turning point MR can be well approximated by a hyperbola (see supplemental materials). Thus, an approximate form for the turning point model, Eq.~\ref{eq:tp}, is
\begin{equation}
    \rho(H) - \rho(0) \approx r_{tp}\sqrt{1+\left(H/H_{tp}\right)^{2}} - r_{tp}.
    \label{eq:tp_a}
\end{equation}
From Fig.~\ref{fig:underdoped}a, we observe that $\rho(0) = \rho_{0} + \alpha' T^{2}$. Plugging in the temperature-dependences, $H_{tp} = \gamma T^{2} + \gamma_{0}$ and $r_{tp} = \beta T^{2} + \beta_{0}$ extracted in Fig.~\ref{fig:underdoped}c, we can rewrite Eq.~\ref{eq:tp_a} with an explicit dependence on temperature. From Fig.~\ref{fig:underdoped}c, we observe that $\gamma_{0}$ and $\beta_{0}$ are much smaller than $\gamma T^{2}$ and $\beta T^{2}$, respectively over the measured range. For example, at the lowest measured temperature 20K, $\beta_{0}/\beta T^{2} \approx 0.3$, and $\gamma_{0} / \gamma T^{2} \approx 0.1$. Thus, the offsets $\gamma_{0}$ and $\beta_{0}$ can be neglected, and we arrive at the approximate scaling relation
\begin{equation}
    \frac{\rho(H) - \rho_{\text{0}}}{T^{2}} \approx \beta \sqrt{1 + \left(\frac{H}{\gamma T^{2}} \right)^{2}} + \alpha' - \beta.
    \label{eq:afmmr}
\end{equation}

\begin{figure}[!htpb]
\centering
\includegraphics[scale=0.8]{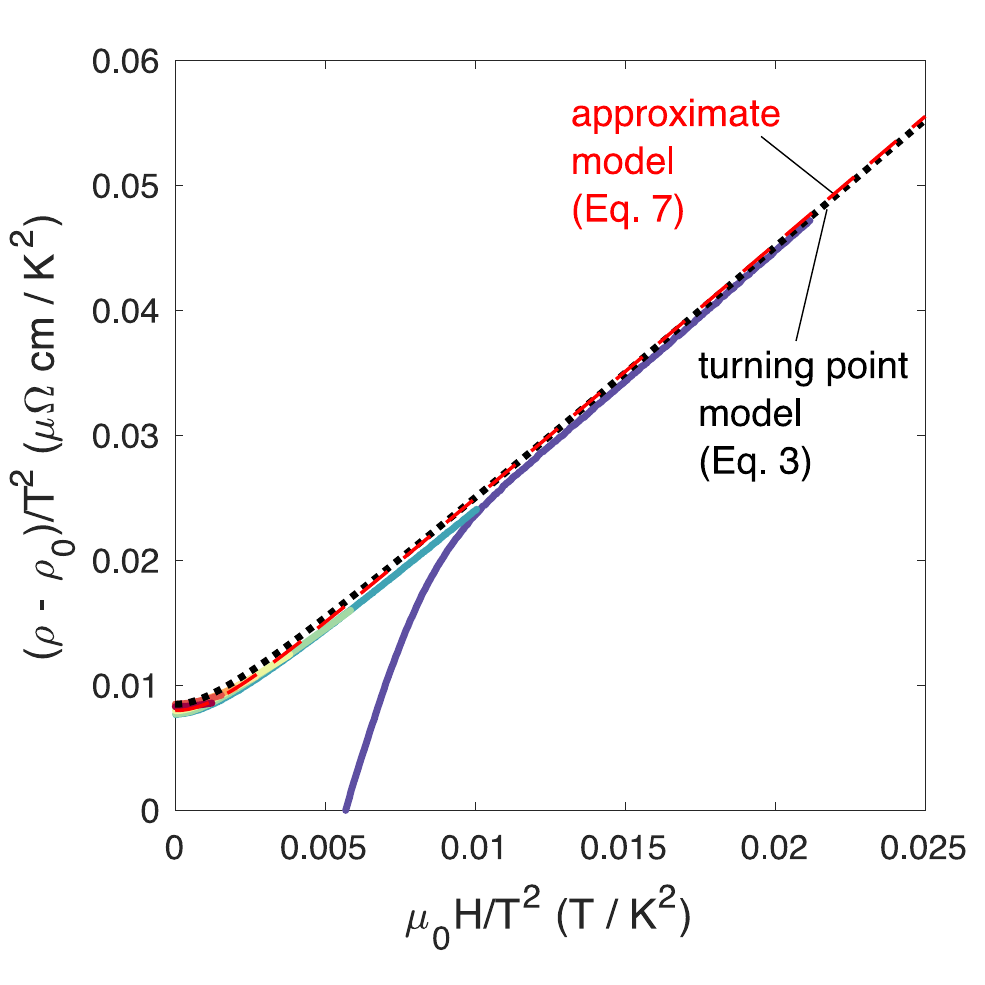}
\caption{{\bf H-T magnetoresistance scaling inside the AFM ordered state of BaFe$_2$(As$_{0.81}$P$_{0.19}$)$_2$} The magnetoresistance data inside the AFM state (blue lines) collapses when the axes are rescaled by temperature squared and the residual resistivity ($\rho_{0} = 122.8~\mu\Omega$cm) is subtracted. The dotted black line is the turning point model given by Eq.~\ref{eq:tp}, where $r_{tp}=0.0035 [\mu\Omega cm / K^{2}] \times T^{2}$ and $H_{tp} = 0.0017 [T/K^{2}] \times T^{2}$. These constants of proportionality are in agreement with the temperature-dependences shown in Fig.~\ref{fig:underdoped}c. The dashed red line is the simplified approximate expression given by Eq.~\ref{eq:afmmr} with $\beta = 0.0039 [\mu\Omega cm / K^{2}]$, and $\alpha' = 0.0085 [\mu\Omega cm / K^{2}$].}
\label{fig:underdoped_scaling}
\end{figure}

Fig.~\ref{fig:underdoped_scaling} shows that both axes can be rescaled by $T^{2}$ to illustrate the applicability of the approximate scaling relation (Eq.~\ref{eq:afmmr}) to the MR data over the measured range. Fig.~\ref{fig:underdoped_scaling} also shows the consistency between the approximate expression given by Eq.~\ref{eq:afmmr}, and the full microscopic turning point model given by Eq.~\ref{eq:tp}. The approximate Eq.~\ref{eq:afmmr} is similar to the established phenomenological $H-T$ scaling relation (Eq.~\ref{eq:qcmr}) in the literature --- the difference is that this in the AFM ordered state rather than the quantum critical regime, and the temperature-dependent resistivity varies with $T^{2}$ rather than with $T$. This is reflected in the different temperature-dependence of the denominator in Eq.~\ref{eq:afmmr} compared to Eq.~\ref{eq:qcmr}.  One of the central findings of the present study is that the $H-T$ MR scaling, previously observed in the context of quantum critical metals~\cite{Hayes2016,Sarkar2019,Prozorov,Licciardello2019,Niu2017,Nakajima2019,Sales2016,Chu2019,Kumar}, is also a feature of the symmetry-broken ordered phase of~\basp.

A key approximation in arriving at Eq.~\ref{eq:afmmr} from the turning point model is that the temperature variation of parameters $H_{tp}$ and $r_{tp}$ is much larger than their zero-temperature values (Fig.~\ref{fig:underdoped}c). In other words, A physical interpretation is that the temperature-dependent scattering rate is anisotropic, and strongly peaked at the turning points due to, for example, inelastic scattering from spin fluctuations~\cite{Tucker2014}. As a result of this mechanism, the inelastic scattering rate at the turning points can be much larger than the background defect scattering rate down to relatively low temperature --- for example, assuming disordering scattering is isotropic, we estimate that the inelastic $T^{2}$ scattering near the turning point is enhanced by  a factor of $[\beta T^{2}/\beta_{0}]/[\alpha' T^{2}/\rho_{0}] \approx 100$ over the background scattering rate at $T = 20K$. Nevertheless, at sufficiently low temperature, it is expected that the elastic disorder scattering contribution becomes important, and the assumptions involved in Eq.~\ref{eq:afmmr} should break down. We expect the $H-T$ scaling shown in Fig.~\ref{fig:underdoped_scaling} to fail when $\beta_{0}/\beta T^{2} \sim 1$ (i.e. at T $<$ 10 K). Unfortunately, the superconducting critical field makes this temperature range inaccessible in our measurement apparatus.

\subsection{Quantum critical regime, \basp~with x = 0.31}
We move now to the quantum critical regime where P-substitution (x = 0.31) suppresses the AFM transition to T = 0, resulting in spin fluctuations that persist to low temperature~\cite{Hu2016,Matan2009,Shibauchi2014}. Motivated by our analysis of the MR in the x = 0.19 sample, we will apply a similar model in this paramagnetic regime. The difference is that, rather than opening a gap at points on the Fermi surface nested by the AFM ordering vector, the AFM coupling causes a strong quasiparticle scattering rate at these points (`hot spots') due to inelastic scattering from spin fluctuations~\cite{Koshelev2016,Richard2011}. Orbital motion across the hot spots provides a linear contribution to the MR~\cite{Koshelev2016,Rosch1999m}.
\begin{equation}
    \rho(H) - \rho(0) = r_{hs}\mathcal{G}\left(H/H_{hs}\right),
    \label{eq:hs}
\end{equation}
where $\mathcal{G}$ is a dimensionless function, with a slightly different exact form compared to the turning point model in Eq.~\ref{eq:tp}, but with qualitatively similar behavior --- $H^{2}$ variation when $H<H_{hs}$ crossing over to $H$-linear when $H>H_{hs}$. We point out that the magnetoresistance is expected to saturate at very large fields $H \gg H_{hs}$ within our model, which is not observed in data even up to $100T$~\cite{Hayes2016}. $H_{hs}$ is a hot spot parameter, which is expected to have the following dependence on temperature and background scattering rate
\begin{equation}
    H_{hs} \propto \frac{T}{\sqrt{\tau_{\text{cold}}}}.
    \label{eq:muhs}
\end{equation}
This parameter quantifies the strength of scattering at the hot spot, and its region of influence as compared to the background scattering rate, $\tau_{\text{cold}}$. Note that increasing $\tau_{\text{cold}}$ can change the `width' of the hot spot. This explicit dependence on background scattering (including isotropic defect scattering) suggests that an effective experimental method to test this model in the quantum critical regime is by varying the concentration of defects in the underlying crystal lattice. We accomplish this with 3 MeV alpha particle irradiation of samples with $x$ = 0.31 phosphorous substitution. This irradiation method produces isotropic defects with a distribution of radii (from point-like to $nm$ in radius)~\cite{Eisterer2018}, which increase the residual resistivity at zero field and temperature. This does not significantly affect the temperature-dependence of the resistivity above the superconducting transition, but does increase the residual resistivity (see supplemental materials), in line with Matthiessen's rule.

\begin{figure*}[!htpb]
\centering
\includegraphics[scale=0.7]{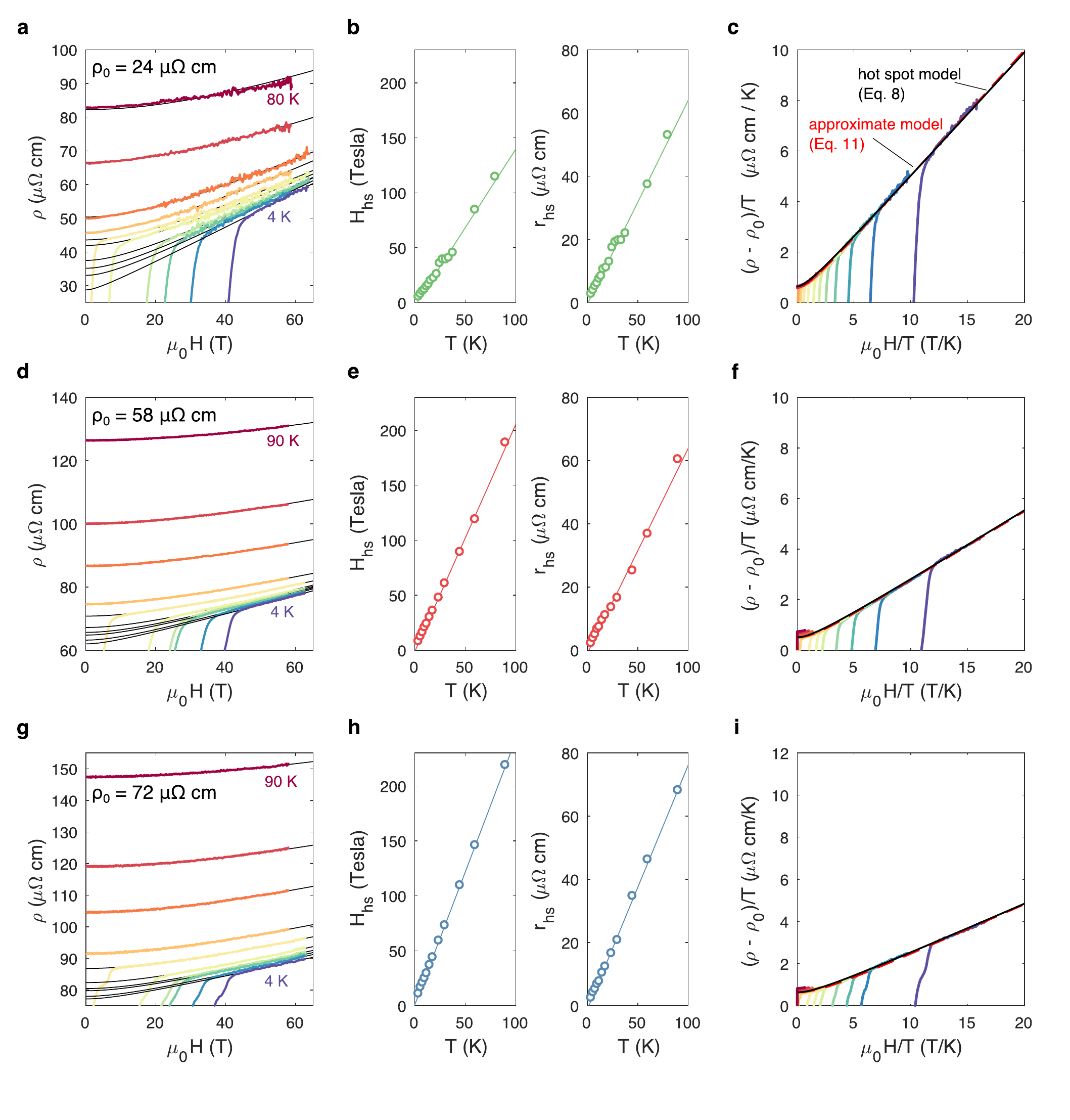}
\caption{{\bf Magnetoresistance in BaFe$_2$(As$_{0.81}$P$_{0.31}$)$_2$ disordered with alpha particle irradiation} (a,d,g) Resistivity versus magnetic field for three different samples with increasing concentration of defects; the residual resistivity ($\rho_{0}$) is shown in each panel. Black lines are fits using the hot spot model described in the main text in Eq.~\ref{eq:hs}. (b,e,h) 
Fit parameters from the hot spot model as a function of temperature for each sample. The slope of $H_{hs}(T) = \gamma T$ increases with defect concentration ((b) $\gamma = $ 1.427 T/K, (e) 2.06 T/K, (h) 2.42 T/K), while $r_{hs}(T)$ is largely unaffected by defect concentration ((b) $\beta = 0.65 \mu\Omega cm/K$, (e) 0.65 $\mu\Omega cm/K$, (h) 0.77 $\mu\Omega cm/K$). (c,f,i) The data in all cases can be rescaled as a function of temperature after subtracting the residual resistivity ($\rho_{0}$). The black lines are Eq.~\ref{eq:afmmr} using the slopes of $r_{hs}(T)$ and $H_{hs}(T)$ in panels e, f \& h, respectively. Dashed red lines are the simplified phenomenological form given by Eq.~\ref{eq:hs}.}
\label{fig:MR}
\end{figure*}

Fig.~\ref{fig:MR} shows the magnetoresistance for samples in the quantum critical regime with varying concentration of defects. The MR data for each sample across a range of temperatures can be well fitted by Eq.~\ref{eq:hs} and Eq.~\ref{eq:muhs}. The temperature-dependent MR data for each sample were fit by assuming that $H_{hs} = \gamma T$; $\gamma$ and $r_{hs}$ are the fit parameters, where $\gamma$ is a constant for each sample, and $r_{hs}$ is extracted independently for each isothermal magnetoresistance trace. The parameters resulting from the fits are shown in Figs.~\ref{fig:MR}b, e, \&~h. Notably, we observe that $\gamma$ increases as the background scattering rate increases (Fig.~\ref{fig:MR}b, e~\& h), in agreement with Eq. \ref{eq:muhs}.

Here, we derive an approximate $H-T$ scaling relation. The experimentally measured resistivity at zero field is given by $\rho(0) = \rho_{0} + \alpha k_{B}T$ (see supplemental). Plugging this into Eq. \ref{eq:hs}, we obtain a simplified expression by using a hyperbolic approximation of the function $\mathcal{G}$
\begin{equation}
    \rho(H) - \rho(0) \approx r_{hs}\sqrt{1 + \left(H/H_{hs} \right)^{2}} - r_{hs}.
    \label{eq:prescaling}
\end{equation}
According to Eq.~\ref{eq:muhs} and Figs.~\ref{fig:MR}c, f \& i, $H_{hs} = \gamma T$, and $r_{hs} = \beta T + \beta_{0}$. The hot spot theory predicts that $H_{hs}(T)$ has a zero-intercept. In addition, we find that $\beta_{0}/\beta T$ is small over this measured temperature range (at 4K, $\beta_{0}/\beta T \approx 0.15$), so we can once again expand in powers of $\beta_{0}/\beta T$ and drop subleading terms. Analogously to Eq.~\ref{eq:afmmr}, the result is
\begin{equation}
    \frac{\rho(H) - \rho_{0}}{T} \approx \beta \sqrt{1 + \left(\frac{ H}{\gamma T} \right)^{2}} - \beta + \alpha.
    \label{eq:qcmr_v2}
\end{equation}
Note that Eq.~\ref{eq:qcmr_v2} has the same form as the $H-T$ scaling form of Eq.~\ref{eq:qcmr}. Figs.~\ref{fig:MR}b, e \& h show the validity of the approximate $H-T$ scaling relation given by Eq.~\ref{eq:qcmr_v2}. For each of these samples $\beta \approx 0.7 \mu\Omega cm / K$ and $\alpha \approx 0.8 \mu\Omega cm / K$.

\begin{figure*}[!htpb]
\centering
\includegraphics[scale=0.75]{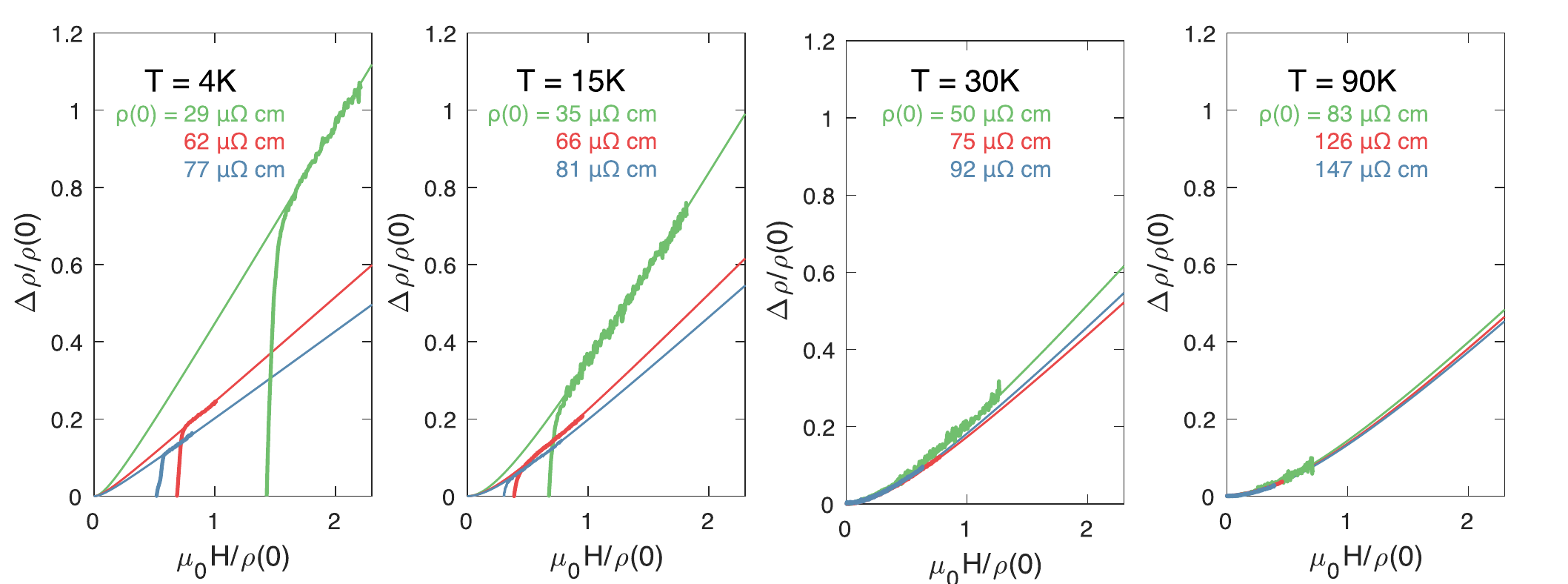}
\caption{{\bf Kohler's rule in quantum critical~\basp~x=0.31} A comparison of the isothermal magnetoresistance of separate samples with varying doses of alpha particle irradiation. Solid lines are fits to Eq.~\ref{eq:hs}, from which the zero-field resistivity is extracted from each trace. The curves are labeled by the resistivity of the sample at zero field at the given temperature. $\mu_{0}H/\rho(0)$ is in units of Tesla/$\mu\Omega$cm.}
\label{fig:kohler}
\end{figure*}

The magnetoresistance in this model has an underlying orbital mechanism, and so in the limit of low magnetic fields one expects a connection to Kohler's rule (Eq.~\ref{eq:kohler}), which is an instrically orbital picture of magnetoresistance governed by a single quasiparticle relaxation time. In the scaling shown in Eq.~\ref{eq:qcmr_v2} as a function of temperature, the resistivity needs to be partitioned into two components --- this is at odds with Kohler's rule. From the perspective of hot spots, Kohler's rule is violated because the relaxation rate is highly anisotropic, and therefore the orbital magnetoresistance cannot be captured by a single timescale as the temperature is varied. An even stricter test of Kohler's rule is at a fixed temperature, where the relaxation rate is uniformly varied by disorder scattering. Even in the case of anisotropic scattering, one expects Kohler's rule to be satisfied at a fixed temperature so long as the `pattern' of anisotropy is unchanged~\cite{sorbello_effects_1975}. In Fig.~\ref{fig:kohler}, we show that at a fixed temperature in~\basp~ ($x=0.31$), Kohler's rule is violated in the high-field linear magnetoresistance regime, but satisfied in the low-field quadratic regime. The failure of Kohler's rule in the linear magnetoresistance regime at a fixed temperature implies that background disorder scattering changes the pattern of scattering anisotropy, for example by changing the size of the hot spot on the Fermi surface. In fact, such an effect is a natural consequence of our hot spot model as discussed in Eq.~\ref{eq:muhs} and the text surrounding it. Moreover, this model also predicts that the effect of hot spots is not pronounced at low fields because the majority of quasiparticles have not yet been pushed into the hot spot. The validity of Kohler's rule in the low-field quadratic regime is a direct validation of this prediction.

\section{Discussion}
The $H$-linear MR of \basp~can be reconciled as an orbital response in a model with a highly anisotropic structure in the Fermi surface --- either due to the presence of a gap or a hot spot --- at points on the Fermi surface nested by the antiferromagnetic ordering vector. This could explain why the $H$-linear MR depends only on the direction of the field and not on the direction of the current~\cite{Hayes2018}, as it is observed only when the cyclotron path crosses the hot spots or turning points. Within this picture, the linear increase of the hot spot scattering strength with temperature underlies the $H-T$ scaling of the magnetoresistance of~\basp~in the quantum critical regime. This requires the approximation that the zero-temperature intercept of $r_{hs}$ (arising from disorder scattering) be treated as a subleading term. Note that this does not imply $\rho_0$ is small compared to $\rho_t$, only that the effect of disorder on the hot spot scattering mechanism is small compared to its temperature dependence, parameterized by $\beta_0/\beta T$. This approximation, and consequently the $H-T$ scaling represented by Eq.~\ref{eq:qcmr_v2}, is expected to break down as $\beta_0/\beta T$ grows with disorder or decreasing temperature. We observe a weak deviation from $H-T$ scaling in the most disordered sample at 1.5 K (see supplemental materials), suggesting that disorder scattering indeed begins to play an important role in the low-temperature limit. However, experiments at lower temperatures and higher fields are necessary to explore this breakdown further. Nevertheless, the validity of Kohler's scaling at low $H/T$ strongly suggests that the magnetoresistance has an orbital origin, and the similarity in the scaling relations between antiferromagnetic and quantum critical compounds implies a common ancestor, in this case the evolution of the turning points into hot spots as the system is doped. 

In this light it is surprising that the coefficients determining the linear dependence on field and temperature respectively are comparable at the quantum critical composition. This appears as a coincidence in the data rather than a direct prediction of our model. There may however, be a more fundamental reason for this connection having to do with the character of spin excitations as a function of doping. Neutron scattering experiments show that the well-defined spin waves of BaFe$_2$As$_2$ become increasingly diffusive spin fluctuations as the material is doped~\cite{Tucker2014}. It is possible that diffusive spin fluctuations are the source of the large inelastic scattering rate centered at hot spots or turning points, which we argue is an important ingredient for $H-T$ MR scaling. In fact, in the parent compound BaFe$_2$As$_2$ where the spin waves are sharply-defined~\cite{Tucker2014}, the $H-T$ scaling seems to break down because $H-$linear MR has a non-universal dependence on disorder~\cite{ishida_manifestations_2011}. Notably, it is thought that the diffusive nature of spin excitations at moderate doping levels also provides a pairing mechanism for superconductivity~\cite{Tucker2014,Wang}, and therefore it would be interesting to explore the possible correlation between $H-T$ MR scaling and superconductivity in~\basp~\cite{Sarkar2019}. The present study shows that MR measurements may be useful for probing hot spot properties across the P-substituted phase diagram, which could provide valuable quantitative information as to how spin excitations influence the resistivity and ultimately superconductivity in iron-based superconductors~\cite{Wang}.

However, the hot spot picture is likely incomplete. For example, the proximity to a nematic quantum critical point~\cite{Kuo2016} suggests nematic fluctuations could also affect the resistivity at zero field~\cite{Lederer2017,Wang,chubukov_origin_2015,fernandes_anisotropic_2011,fernandes_preemptive_2012}. Fe(Se$_{1-x}$S$_{x}$) has a nematic quantum critical point with some evidence of magnetic fluctuations~\cite{baum_frustrated_2019,gati_bulk_2019}, and shows very similar magnetoresistance behavior to that of~\basp~\cite{Licciardello2019}. Interestingly, in Fe(Se$_{1-x}$S$_{x}$) the conventional MR from the cold spots adds a parallel contribution to the conductivity which can be suppressed with disorder to restore scaling behavior of Eq.~\ref{eq:qcmr}~\cite{Licciardello2019}. It would be interesting to extend the present antiferromagnetic hot spot picture to the case of nematic fluctuations.

Just as the scaling observed in the antiferromagnetic compositions arises because the turning points dominate both the MR and the temperature-dependent scattering rate, so too does the scaling observed at quantum critical compositions suggest hot spots dominate the two phenomena. While the hot spot regions are expected to give a correction to the conductivity that is linear in $T$ (see supplemental), a question remains as to why this correction is not shorted out by the regions of the Fermi surface away from the hot spots, where the scattering rate is comparatively low ~\cite{Hlubina1995}. Within a nearly antiferromagnetic Fermi liquid framework, one option is that mixing of hot spot and disorder scattering leads to nontrivial behavior of the overall resistivity~\cite{Rosch1999,Rosch1999m}. Our data is difficult to reconcile with such a mechanism for temperatures above $T_{c}$ --- disorder adds a temperature independent component to the resistivity consistent with Matthiessen's rule, which can only occur if the temperature-dependent scattering is independent of disorder. Other nearly antiferromagnetic Fermi liquid models show that magnetic fluctuations at hot spots can influence the overall resistivity through multiple scattering~\cite{Breitkreiz2014}, or the so-called backflow effect~\cite{Kontani2008}. A more recent revival of the hot spot picture outlined by the theory of Mousatov, Hartnoll and Berg as applied to Sr$_3$Ru$_2$O$_7$, has shown that an unconventional two-particle scattering process connecting hot and cold regions can render the entire Fermi surface a `marginal' Fermi liquid with an overall $T$-linear resistivity~\cite{Berg2018}. For the present model to remain valid, the magnetotransport would still have to originate from orbital dynamics across hot spots even in the marginal Fermi liquid picture.

Finally, hot spots at the antinodal regions of the Brillouin zone of the cuprates have been suggested as a source of anomalous behavior in transport and photoemission measurements for some time~\cite{Armitage2002,Kontani1999}, so in light of our results the recent observations of linear-in-$H$ magnetoresistance in both electron and hole doped cuprate superconductors is perhaps unsurprising~\cite{Sarkar2019,Giraldo-Gallo2018}. This picture of an anisotropic scattering rate would also explain a long-standing question regarding the violation of Kohler's rule observed in the cuprates~\cite{Harris1995}. Moreover, scaling behavior with the cotangent of the Hall angle has  been observed, the so-called `modified' Kohler's rule~\cite{Kontani2002}, and we leave to future work whether this can similarly be explained by the present hot spot model.

\section{Acknowledgements}
We thank F.F. Balakirev, and R. McDonald for their support during pulsed field experiments. We also thank R. McDonald, A. Patel, E. Altman, E. Berg, R. Fernandez, D. Maslov, and N.E. Hussey for valuable discussions and insights. Experimental data on pristine BaFe$_2$(As$_{1-x}$P$_x$)$_2$ (x = 0.31) was previously published in Ref.~\cite{Hayes2016}, and reprinted here with permission. This work was supported as part of the Center for Novel Pathways to  Quantum  Coherence in  Materials, an Energy Frontier Research Center funded by the U.S. Department of Energy, Office of Science. N.M. was  supported by the Gordon and Betty Moore Foundations EPiQS Initiative through Grant GBMF4374. Work of A.E.K. in Argonne was supported by the U.S. Department of Energy, Office of Science, Basic Energy Sciences, Materials Sciences and Engineering Division. A portion of this work was performed at the National High Magnetic Field Laboratory, which is supported by the National Science Foundation Cooperative Agreement No. DMR-1157490 and the State of Florida.

\section{Author Contributions}
N.M. fabricated the devices and carried out the low-field measurements. N.M., Y.L., and T.S. carried out sample irradiation. N.M., V.N., I.M.H., and J.S. carried out the pulsed field experiments. N.M. and A.E.K. performed the theoretical fitting. I.M.H. grew the crystals. All authors contributed to writing the manuscript.

\pagebreak
%

\end{document}